\documentclass[prc,aps,nofootinbib,showkeys,showpacs,twocolumn]{revtex4} 

\usepackage{amsmath}
\usepackage{amssymb}
\usepackage{amsfonts}
\usepackage{color}

\usepackage{epsfig}
\usepackage{graphicx}
\begin{document}
\date{\today}

\title{Density Matrix Functional Theory for the Lipkin model
}

\author{Denis Lacroix.} \email{lacroix@ganil.fr}
\affiliation{GANIL, CEA and IN2P3, Bo\^ite Postale 5027, 14076 Caen Cedex, France}

\begin{abstract}  
A Density Matrix Functional theory is constructed semi-empirically for the two-level 
Lipkin model. This theory, based on natural orbitals and occupation numbers, 
is shown to provide a good description for the ground state energy of the system 
as the two-body interaction and particle number vary. The application of Density Matrix Functional theory  
to the Lipkin model illustrates that it could be a valuable tool for systems presenting a shape phase-transition such as 
nuclei.  
The improvement of one-body 
observables  description as well as the interest for Energy Density Functional theory are discussed.
\end{abstract}


\maketitle

\section{Introduction}

Recently, large efforts are devoted to the construction of an Energy Density 
Functional (EDF) able to described at best properties of nuclei over the whole
nuclear chart\cite{Ben03,Sto07}. The standard strategy to design an EDF for nuclei
is to start with a single-reference EDF (SR-EDF) where an effective interaction (Skyrme or Gogny type)
and a trial state (Slater Determinant or more generally quasi-particle vacuum) are chosen.
This technique is able to describe short-range correlations like pairing and provides already
a rather good description of observables such as masses, under the condition that some symmetries of the 
original Hamiltonian are broken. The SR-EDF is then extended to restore broken symmetries and/or incorporate 
long-range correlations through configuration mixing, 
leading to the so-called Multi-Reference EDF (MR-EDF)\cite{Rin80}.  Recent applications of this 
technique have revealed important conceptual and practical difficulties \cite{Dob07}
related to the absence of a constructive framework 
for multi-reference calculations. A solution to this problem,
has been recently proposed \cite{Lac08} and successfully tested in nuclei \cite{Ben08}. However, this cure 
does not apply  to most of the functionals currently used \cite{Dug08}, i.e. those with fractional powers of density.

This motivates the search of new techniques to extend actual SR-EDF. The Density Matrix Functional Theory (DMFT)
\cite{Gil75} appear as an alternative to configuration mixing \cite{Umr00}. Although this theory was proposed more 
than 30 years ago \cite{Gil75}, explicit forms of functionals and applications have only been explored rather recently. 
There are nowadays an increasing interest in proposing accurate DMFT \cite{Kol06}. In this work, DMFT 
is applied to the two-level Lipkin model \cite{Lip65}. In this model, the Hartree-Fock (HF) theory fails to reproduce 
the ground state energy \cite{Aga66} while configuration mixing like Generator Coordinate Method (GCM) provides
a suitable tool \cite{Rin80}. Therefore, the two-level Lipkin model is perfectly suited both to illustrate that 
DMFT could be a valuable tool and to provide an example of functional for system with a 
"shape" phase-transition. 

In this following, some aspects of DMFT are first recalled. Then a semi-empirical 
functional is constructed for the Lipkin model and applied for various particle number and two-body interaction strengths.
It is shown to improve significantly the HF theory. Finally, the interest of constructing more general 
functional of natural orbitals and occupation numbers in the EDF context is outlined. 
       
\section{Discussion on DMFT}

The concept of Density Matrix Functional Theory is a generalization of the Hohenberg-Kohn theorem \cite{Hoh64} 
due to Gilbert \cite{Gil75}. It relies on a theorem showing that 
the ground state energy could be written as a functional of the one-body density 
matrix (OBDM) $\gamma(\mathbf{r},\mathbf{r'})$ (instead of the local one-particle density $\rho(\mathbf{r}) 
\equiv \gamma(\mathbf{r},\mathbf{r})$ in the standard Hohenberg-Kohn theorem). Then, similarly to the Kohn-Sham orbitals \cite{Koh65}, the 
eigenvalues $n_i$ and eigenvectors $\varphi_i$, called hereafter resp. {occupation numbers} and natural orbitals, of the OBDM are often used instead of $\gamma(\mathbf{r},\mathbf{r'})$ with the relation $\gamma = \sum_i | \varphi_i \rangle n_i \langle \varphi_i |$. 
The variation of the functional     
\begin{eqnarray}
{\cal F}[\{\varphi_i \}, \{n_i \}] &=& {\cal E} [\{\varphi_i \}, \{n_i \} ] \nonumber \\
&-&\mu \{ Tr(\rho) -N \} -\sum_{ij} \lambda_{ij} (\langle \varphi_i | 
\varphi_j \rangle  - \delta_{ij}) ,
\label{eq:dmft}
\end{eqnarray} 
with respect to one-particle state components $\varphi_i^*(\mathbf{r})$ and occupation numbers (with the additional constraint  $0 < n_i < 1$) is then 
performed to obtained the optimal $\varphi_i$, $n_i$ and associated ground state energy.
The set of Lagrange multipliers $\mu$ and $\{ \lambda_{ij} \}$ are introduced to insure particle number conservation and 
orthogonality of the single-particle states. In Eq. (\ref{eq:dmft}), ${\cal E} [\{\varphi_i \}, \{n_i \} ]$ is nothing but the functional itself which has to be found. In electronic system, the functional is generally separated into the Hartree, denoted by ${\cal E}_{H}$, 
(eventually Hartree-Fock, ${\cal E}_{HF}$) part and the exchange-correlation part, denoted here by ${\cal E}_{XC}$ (eventually 
correlation only ${\cal E}_C$).

While the DMFT has been studied theoretically for a rather long time \cite{Gil75,Val80a,Val80b,Zum85,Mul84}, only recently explicit functionals of the OBDM or directly to natural orbitals have been proposed and applied 
to realistic situations\cite{Goe98,Csa00,Csa02,Yas02,Kol04,Cio03,Per04,Gri05,Lat05,Cio05,Lei05,Kol06,Mar08,Lat08}. 
There are nowadays extensive works to test  functionals especially in infinite systems, the so-called 
Homogeneous Electronic Gas (HEG) \cite{Cio99,Lat07}.  
     
  
\section{Application of the DMFT to the Lipkin model}

The "Lipkin Model" \cite{Lip65} is an exactly 
solvable model that has often been used as a benchmark for
approximations for the nuclear many-body problem \cite{Rin80}.
In this model, one considers $N$ particles distributed in two N-fold degenerated shells separated 
by an energy $\varepsilon$. The associated Hamiltonian is given by:
\begin{eqnarray}
H = \varepsilon J_0 - \frac{V}{2} (J_+ J_+ + J_- J_-) , 
\label{eq:hamillipkin}
\end{eqnarray}
where $V$ denotes the interaction strength while $J_0$, $J_\pm$ are the quasi-spin operators defined as
\begin{eqnarray} 
J_0 &=& \frac{1}{2} \sum_{p=1}^{N} \left(c^\dagger_{+,p}c_{+,p} - c^\dagger_{-,p}c_{-,p}\right) , \nonumber \\
J_+ &=& \sum_{p=1}^{N} c^\dagger_{+,p}c_{-,p},~~~ J_- = J_+^\dagger ,\nonumber
\end{eqnarray} 
$c^\dagger_{+,p}$ and $c^\dagger_{-,p}$ are creation operators associated with the upper and lower level.
The exact solution of this model, is easily obtained by noting that $J^2$ (but not $J_0$) commute with $H$. It is 
then convenient to introduce the basis of eigenstates of $J^2$ and $J_0$ and diagonalize the Hamiltonian in this 
particular space (for more detail see for instance \cite{Sev06}). 

\subsection{Hartree-Fock approximation}

In the Hartree-Fock (or Mean-Field) theory, the many-body wave function is replaced by a Slater Determinant (SD) given by $| \Phi 
\rangle = \Pi_{p=1}^N a^\dagger_{0,p} | - \rangle$. Here, a new single-particle basis, denoted by 
$\{\varphi_{0,p},\varphi_{1,p} \}$ associated to the set of creation/annihilation 
operators $\{a^\dagger_{0,p},a^\dagger_{1,p} \}$ has been introduced through the relation  
\begin{eqnarray}
\left(
\begin{array} {c}
a^\dagger_{1,p} \\
a^\dagger_{0,p}
\end{array}
\right) &=& 
\left( 
\begin{array} {cc}
f^* & -g^* \\
g & f  
\end{array}
\right)
\left(
\begin{array} {c}
c^\dagger_{+,p} \\
c^\dagger_{-,p}
\end{array}
\right),
\label{eq:matac}
\end{eqnarray}
where the choice 
\begin{eqnarray}
f = \cos(\alpha), ~~~g = \sin(\alpha) e^{i\varphi} ,
\end{eqnarray}   
automatically insures the orthogonality of the new states. Due to simple structure of the Lipkin model, 
the variation with respect to the SD state is identical to the variation of the $\alpha$ and $\varphi$, i.e. 
$| \Phi \rangle = | \Phi (\alpha , \varphi) \rangle$ and the HF energy becomes a functional 
of these parameters:    
\begin{eqnarray}
{\cal E}_{MF}(\alpha,\varphi) \equiv \left\langle \Phi(\alpha, \varphi) | H | \Phi (\alpha, \varphi)  \right\rangle 
\end{eqnarray}
Anticipating for the forthcoming discussion, we first write ${\cal E}_{MF}$ as a functional 
of the  OBDM $\gamma$:
\begin{eqnarray}
{\cal E}_{MF}[\gamma] &=& \varepsilon {\rm Tr}(J_0 \gamma) \nonumber \\
&-& \frac{V(N-1)}{2 N} \Big\{ ({\rm Tr}[\gamma J_+ ])^2 + ({\rm Tr}[\gamma J_-])^2 \Big\}  .
\label{eq:dmft1lipkin}
\end{eqnarray}
In the Hartree-Fock limit, the OBDM contains all the information on the many-body state and 
simply reads $\gamma=\sum_{p=1}^N | \varphi_{0,p} \rangle \langle \varphi_{0,p} |$. Reporting the expression 
of $| \varphi_{0,p} \rangle$ in terms of $\alpha$ and $\varphi$, we recover the standard HF expression:
\begin{eqnarray}
{\cal E}_{MF}[\alpha , \varphi] &=& 
-\frac{\varepsilon N}{2} \left\{ \cos(2\alpha) + \frac{\chi}{2} \sin^2(2\alpha) \cos(2\varphi) \right\}.
\label{eq:hflipkin}
\end{eqnarray}
where $\chi = V(N-1) / \varepsilon$. Minimizing with respect to $(\alpha,\varphi)$ leads to HF energy, denoted by 
${\cal E}^0_{HF}$ (both with $\varphi=0$):
\begin{eqnarray}
{\cal E}^0_{HF} &=& - \frac{\varepsilon N}{2}  ~~{\rm for} ~~\chi \leq 1 ~~~( {\rm at}~ \alpha =0) , \nonumber \\
{\cal E}^0_{HF} &=& - \frac{\varepsilon N} {4\chi} \left( 1+ \chi^2 \right) \nonumber 
~{\rm for} ~\chi > 1 ~( {\rm at}~ \chi\cos(2\alpha) = 1) .
\end{eqnarray}
The HF solution for the Lipkin model has been extensively discussed in the 
literature \cite{Aga66,Rin80}. While it provides a rather good estimate of the exact 
energy in the weak coupling or in the large $N$ limit, it generally differs rather 
significantly from it in particular for $\chi \sim 1$. This  discrepancy 
essentially reflects the failure of the HF method to account for configuration 
mixing in a single-reference framework for systems with a shape phase-transition \cite{Rin80}. 

\subsection{Expression of the Hamiltonian for a General correlated state}

Due to the two-body nature of the Hamiltonian (Eq. (\ref{eq:hamillipkin})), 
the most natural way to extend the Mean-Field framework to correlated system is to introduce 
the two-body density matrix, denoted by $\Gamma_{12}$ and the associated correlation 
matrix $\sigma_{12}$ (see for instance \cite{Cio00}). 
Using the OBDM of the correlated system $\gamma$, $\sigma_{12}$ is defined through the relation:
\begin{eqnarray}
\Gamma_{12} = \gamma_{1} \gamma_{2} (1-P_{12}) + \sigma_{12} ,
\end{eqnarray} 
where $P_{12}$ denotes the anti-symmetrization operator while the label "i" in $\gamma_i$ refers to the particle on which the density is applied, i.e. $\langle ij |\gamma_{1} \gamma_{2}| kl  \rangle = \langle i|\gamma| k  \rangle \langle j |\gamma| l  \rangle$ \cite{Lac04}. 
The expectation value of the energy then splits into a mean-field part and a correlated part as :
\begin{eqnarray}
{\cal E} = {\cal E}_{MF} [\gamma ] +  {\cal E}_{C} [\sigma_{12}],
\label{eq:emfec0}
\end{eqnarray} 
where ${\cal E}_{MF} [\gamma ]$ is given by expression (\ref{eq:dmft1lipkin}) except that $\gamma$ now refers to the OBDM of the 
correlated system. 
${\cal E}_{C}$ reads:
\begin{eqnarray}
{\cal E}_{C} [\sigma_{12}] &=& -\frac{V}{2} {\rm Tr} \left\{ [J_+J_+ + J_- J_-]_{12} \sigma_{12} \right\} \nonumber \\
&=& -V \sum_{p,p'} \Re\left( \left\langle  +, p ; + ,p' ~ |\sigma_{12} | - , p ; - , p' \right\rangle \right) ,
\end{eqnarray}  
where the notation $[.]_{12}$ emphasizes that the trace is performed using the two-body matrix elements of the operators. 

Eq. (\ref{eq:emfec0}) is valid for any correlated system including the exact ground state. The minimization of it 
with respect to any kind of OBDM and correlations will therefore lead to the ground state energy. 
Such a direct minimization is complex due to the number of degrees of freedom 
involved in the variation \cite{Cio00}.

DMFT provides a practical solution to this problem. Indeed, according to this theory \cite{Gil75}, the total energy could be written as
a functional of $\gamma$. Since ${\cal E}_{MF}[\gamma]$ is already written as a functional of the OBDM, we are left to find a 
functional for ${\cal E}_{C}$. In many cases, both ${\cal E}_{MF}$ and ${\cal E}_{C}$ are directly written as a functional 
of the natural orbitals $\varphi_i$ and occupation numbers $n_i$, i.e.
\begin{eqnarray}
{\cal E}[\{\varphi_i , n_i\}] = {\cal E}_{MF} [\{\varphi_i , n_i\}] +  {\cal E}_{C} [\{\varphi_i , n_i\}] .
\label{eq:emfec}
\end{eqnarray}
DMFT has some additional advantages compared to standard DFT. Besides the fact that the exchange contribution could be expressed 
exactly with the OBDM, at the minimum the optimal OBDM identifies with 
the ground state OBDM. Therefore, expectation values of any one-body observable could be computed and should correspond to 
the ground state expectation value. Accordingly, if the mean-field prescription is used in ${\cal E}_{MF}$ for a given $H$, 
the value of ${\cal E}_{MF}$ at the minimum will truly correspond to the true mean-field contribution to the total 
energy. Consequently, ${\cal E}_{C}$ will truly correspond to the contribution of correlation (this issue will further 
be discussed in section \ref{sec:obobs}).     

Recently, significant efforts have been made in the practical 
construction of such density matrix functionals
\cite{Goe98,Csa00,Csa02,Yas02,Kol04,Cio03,Per04,Gri05,Lat05,Cio05,Lei05,Kol06,Mar08,Lat08}. In most cases, guided by general considerations \cite{Mul84} and/or 
constraints on the two-body density \cite{Csa00}, $\sigma_{12}$ is first written as 
a functional of the natural orbitals and occupation numbers. This corresponds generally to a starting point for more elaborated functionals.
Functionals are further enriched by incorporating additional terms either to correct for the self-interaction problem 
or eventually to better reproduce some specific limits in the infinite HEG or configuration interaction 
calculation in molecules (for a recent review see \cite{Klo07}).

\subsection{Construction of a density matrix functional for Correlated state}
\label{sec:semi}  
Due to the specific form of the Lipkin Hamiltonian given by Eq. (\ref{eq:hamillipkin}), 
$\gamma$ simply writes in the natural basis as:
\begin{eqnarray}
\gamma = \sum_{p=1}^N \Big\{| \varphi_{0,p} \rangle n_0 \langle \varphi_{0,p} | + 
| \varphi_{1,p} \rangle n_1 \langle \varphi_{1,p} | \Big\} ,
\label{eq:obdm}
\end{eqnarray}
with $n_1 = (1-n_0)$. The single-particle states $| \varphi_{ i,p} \rangle$ (with $i=0,1$) now stand for the natural orbitals while the $n_i$ 
correspond to occupation numbers. Similarly to the HF theory, creation operators, 
denoted by $a^\dagger_{i,p}$, associated to natural orbitals are
expressed from the $c^\dagger_{\pm,p}$ using Eq. (\ref{eq:matac}).
The mean-field contribution is easily deduced from the HF case, 
using Eq. (\ref{eq:dmft1lipkin}) and the expression of $\gamma$ given above, leads to:
\begin{widetext}
\begin{eqnarray}
{\cal E}_{MF}(\{ \varphi_{i,p}, n_i \}) & = &  {\cal E}_{MF}(\alpha,\varphi,n_0) \nonumber \\
&=& -\frac{\varepsilon}{2} N  \Big\{ \cos(2\alpha) (2n_0 -1) 
+ \frac{\chi}{2} \sin^2(2\alpha) \cos(2\varphi) (2n_0 -1)^2 \Big\} .
\label{eq:emflipkin}
\end{eqnarray}
\end{widetext}

Expressing ${\cal E}_C$ is less straightforward. 
A possible strategy to construct 
functionals is to identify specific limits at which explicit forms are known.
For instance the two electron case which has been studied in \cite{Low56} has largely influence 
presently used DMFT for molecules \cite{Klo07}. Following this idea, the $N=2$ case is first considered.    
 
\subsubsection{The $N=2$ case}

\begin{figure}[t!]
\includegraphics[height=5.cm]{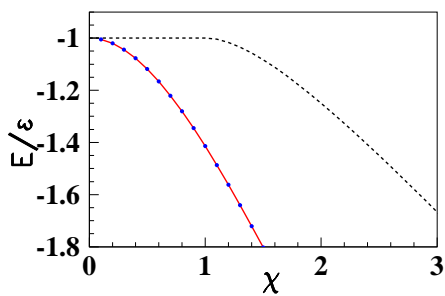}
\caption{Comparison between the exact energy (solid line), Hartree-Fock (dashed line) and  the energy obtained 
from the minimization of Eq. (\ref{eq:funcn2}) (filled circles) 
as a function of $\chi$ for the $N=2$ case.} 
\label{fig:N2}
\end{figure}

To study the $N = 2$ case, a basis of Slater Determinant is constructed using the natural orbitals, i.e. 
$| \Phi_{ij} \rangle = a^\dagger_{i,p}a^\dagger_{j,p'}| - \rangle$ with $i$ and $j$ either $0$ or $1$. The 
ground state wave-function $\Psi$ then reads:
\begin{eqnarray}
| \Psi \rangle = \sum_{ij} C_{ij} | \Phi_{ij} \rangle .
\end{eqnarray}
From the expression of the OBDM and using the fact, that the single-particle states are natural orbitals we obtain the simple relation $|C_{ij}|^2 = \delta_{ij} n_{i}$. From which we deduce $C_{ij} = \delta_{ij} e^{i\theta_{ii}} \sqrt{n_i}$.  As illustrated below, the simplest choice $e^{i\theta_{ii}}=1$  is convenient and leads to
\begin{eqnarray}
| \Psi \rangle = \sqrt{n_0} | \Phi_{00} \rangle +  \sqrt{n_1} | \Phi_{11} \rangle ,
\label{eq:phin2}
\end{eqnarray}
which is nothing but the exact ground state wave-function written as a functional of occupation numbers and natural states. Reporting
this expression in $\left\langle \Phi | H |\Phi \right\rangle$, leads to a total ground 
state energy ${\cal E}^{^{{N=2}}}$ given by:
\begin{widetext}
\begin{eqnarray}
{\cal E}^{^{{N=2}}}(\alpha, \varphi, n_0) &=&  -\varepsilon 
\cos(2\alpha) (2n_0 -1)   
- V \Big\{ \frac{1}{2} \sin^2(2\alpha) \cos(2\varphi) + 2\left( \sin^4(\alpha) \cos(4 \varphi)
+ \cos^4(\alpha) \right)  \sqrt{n_0 (1-n_0)} \Big\} .
\label{eq:funcn2}
\end{eqnarray} 
Using the expression of the mean-field contribution (Eq. (\ref{eq:emflipkin})), we deduce 
\begin{eqnarray}
{\cal E}^{^{{N=2}}}_C(\alpha, \varphi, n_0) &=& - 2 V \Big\{\sin^2(2\alpha) \cos(2\varphi) n_0(1-n_0) 
+\left(\sin^4(\alpha) \cos(4 \varphi)+ \cos^4(\alpha)\right) \sqrt{n_0 (1-n_0)} \Big\}.
\end{eqnarray}
\end{widetext}
Since the above functional is exact, the ground state energy should be recovered by minimization with respect to $n_0$, 
$\alpha$ and $\varphi$. As in the HF case, $\varphi=0$ could always be taken. The variation of $n_0$
should be made under the constraint $n_0 \in [0,1]$. A similar technique as in the HF+BCS case could be employed \cite{Rin80}. 
Writing $n_0 = \cos^2(\theta)$ with $\theta \in [0,\pi/2]$, gives at the minimum 
\begin{eqnarray}
\tan(2\theta) =  \chi \left(\frac{ \sin^4(\alpha) \cos(4 \varphi)+ \cos^4(\alpha)}{\cos(2\alpha)} \right).
\end{eqnarray}  
Then, variation of the $\alpha$ is made to obtain the minimum energy. The result (filled circles) is displayed in Fig. \ref{fig:N2}
and compared to the exact ground state energy given by $E=-\sqrt{1+\chi^2}$ \cite{Lip65} (solid line) and
the Hartree-Fock energy (dashed line). Not surprisingly, while the HF curve deviates significantly from the solid line,
the DMFT could not be distinguished from the exact case. 

\subsubsection{DMFT for $N \ge 3$ and large N scaling}

The Lipkin model with $N=2$ is an interesting pedagogical example of DMFT where the exact 
energy functional in terms of natural orbitals and occupation numbers is known. 
This limit is used here as a guide to provide a DMFT for $N \ge 3$. 
The most simple extension of the functional 
derived for $N=2$, consists in assuming that the interaction of the $N$ particle could be written as a sum of the 
interaction of the $N(N-1)/2$ pairs of particles, 
each pair contributing to the total energy as in the $N=2$ case. This prescription naturally leads to the 
mean-field Hamiltonian given (\ref{eq:emflipkin}) and amount to take for all $N$   
\begin{eqnarray}
{\cal E}_C(\alpha, \varphi, n_0) =  \frac{N(N-1)}{2} {\cal E}^{^{{N=2}}}_C(\alpha, \varphi, n_0).
\label{eq:corsimp}
\end{eqnarray}
\begin{figure}[t!]
\includegraphics[height=9.cm]{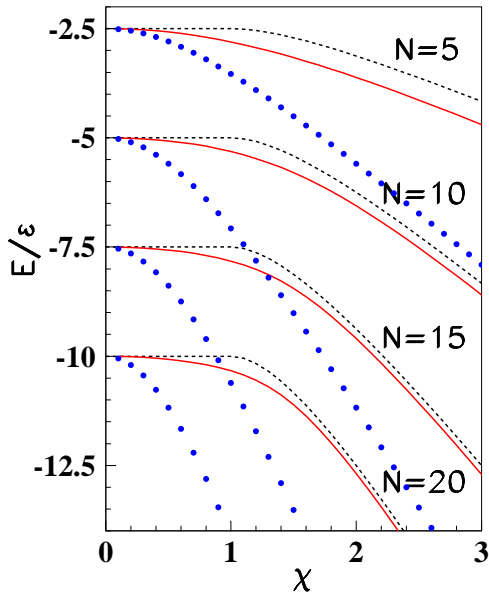}
\caption{Exact ground state energy (solid lines) displayed 
as a function of $\chi$ for $N=5$ to $20$ resp. from top to bottom.
In each case, the corresponding HF (dashed line) and DMFT (filled circles) are shown. The latter are 
obtained by minimization of the functional using the mean-field and correlation energy resp.
given by Eq. (\ref{eq:emflipkin}) and Eq. (\ref{eq:corsimp}).} 
\label{fig:chiref}
\end{figure}
The minimal total energy obtained by varying both occupation numbers and $\alpha$ (using $\varphi=0$) using this
prescription is displayed in Fig \ref{fig:chiref}  (solid circles) as a function of $\chi$ for various particle numbers.
In each case, the exact solution (solid line) and the Hartree-Fock prescription (dashed line) are shown. 
The simple scheme using (\ref{eq:corsimp}) clearly provides a very poor approximation for the ground state energy and
always leads to an energy much below the exact one. In addition, the discrepancy increases as $N$ increases.
This failure points out the complex many-body correlations present in the Lipkin model coming from  
the mixing of 1 particle-1 hole (1p-1h), 2p-2h, $n$p-$n$h excitations in the component in the ground state. This leads to a much more 
complex situation than the $N=2$ case (Eq. (\ref{eq:phin2})). In particular, using the contracted Schr\"odinger 
equation (CSE), we do expect that the two-body correlation depends on higher order correlation 
matrices \cite{Yas02}. The correlation energy given by (\ref{eq:corsimp}) clearly neglects these higher-order effects.
 
From Fig. \ref{fig:chiref}, we see that the correlation energy is largely overestimated. In the following, we use a slightly 
different prescription for the correlation energy given by:      
\begin{eqnarray}
{\cal E}^{^{{N\ge 3}}}_C(\alpha, \varphi, n_0) =  \eta(N) \frac{N(N-1)}{2} {\cal E}^{^{{N=2}}}_C(\alpha, \varphi, n_0) ,
\label{eq:coreta}
\end{eqnarray}
where $\eta(N)$ is a reduction factor  introduced to mimic the effect of higher order correlations.
The optimal value of $\eta$ this is determined empirically from the following procedure. For a given 
value of $N$ and $\eta$, the mimimum energy of the corresponding DMFT, 
denoted by ${\cal E}^N_{min} (\eta, \chi)$  is computed for $\chi$ between 0 and 
$\chi_{max} = 3$. Then, the quantity $D^N(\eta)$, given by
\begin{eqnarray}
D^N (\eta) = \int_0^{\chi_{max}} \left \{ {\cal E}^N_{GS}(\chi) - {\cal E}^N_{min} (\eta, \chi)\right\}^2 d\chi ,  
\end{eqnarray}
where ${\cal E}^N_{GS}(\chi)$ denotes the exact ground state energy for a given $N$ and $\chi$, is computed.
Obviously, $D^N (\eta)$ gives a measure of the deviation between the DFMT minimum energy and the exact energy
over the interval $\chi \in [0,\chi_{max}]$. The optimal value of $\eta(N)$ is defined as the minimum of  $D^N (\eta)$ 
as $\eta$ varies between 0 (the HF limit) and $1$ (the prescription of Eq. (\ref{eq:corsimp})).
\begin{figure}[t!]
\includegraphics[height=5.cm]{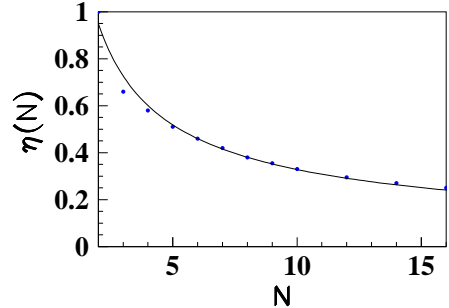}
\caption{Values of the optimal quenching factor $\eta(N)$ as a function of the particle number N. The solid line 
represents the function $\eta(N) = c~N^{-2/3}$ with $c=1.5$.} 
\label{fig:fit}
\end{figure}
Values of optimal 
reduction factors are reported by filled circles in Fig. \ref{fig:fit} as a function of $N$. 
As guessed from the increasing 
discrepancy observed with increasing $N$ observed in Fig. \ref{fig:chiref}, the higher is $N$ the 
smaller should $\eta$ be taken. The variation of $\eta(N)$ turns out to simply behave as $N^{-2/3}$. 
The solid line in Fig (\ref{fig:fit}) represents the function 
\begin{eqnarray}
\eta(N) = c ~ N^{-2/3},
\label{eq:etan}
\end{eqnarray}  
with $c=1.5$ deduced by fitting optimal values of $\eta$. In the following, we show that the 
semi-empirical density matrix functional theory constructed from  the 
mean-field and correlation energy resp. given by Eq. (\ref{eq:emflipkin}) and Eq. (\ref{eq:coreta})
significantly improves the HF theory.

Combining expression (\ref{eq:etan}) with Eq. (\ref{eq:coreta}) shows that the correlation energy scales 
as ${\cal E}_C \propto N^{4/3}$ as $N$ increases. Since this correlation energy is proportional to 
$\langle J^2_x \rangle$, we observe that the $N^{-2/3}$ deduced empirically is nothing but the large $N$ scaling 
behavior obtained analytically in ref. \cite{Dus04}, i.e. $\langle J^2_x \rangle/N^2 \propto N^{-2/3}$.   

\subsection{Results of the semi-empirical DMFT}
 
\begin{figure}[t!]
\includegraphics[height=9.cm]{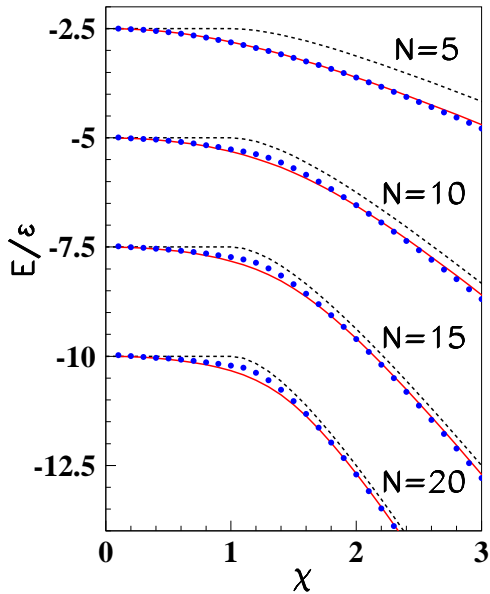}
\caption{Exact ground state energy (solid lines) displayed 
as a function of $\chi$ for $N=5$ to $20$ resp. from top to bottom.
In each case, the corresponding HF (dashed line) and DMFT (filled circle) minimum energy are shown.
The DMFT calculation is performed using the mean-field and correlation energy resp. given by 
Eq. (\ref{eq:emflipkin}) and Eq. (\ref{eq:coreta}) with $\eta(N) =1.5 ~{N}^{-2/3}$.} 
\label{fig:chi}
\end{figure}

\begin{figure}[t!]
\includegraphics[height=9.cm]{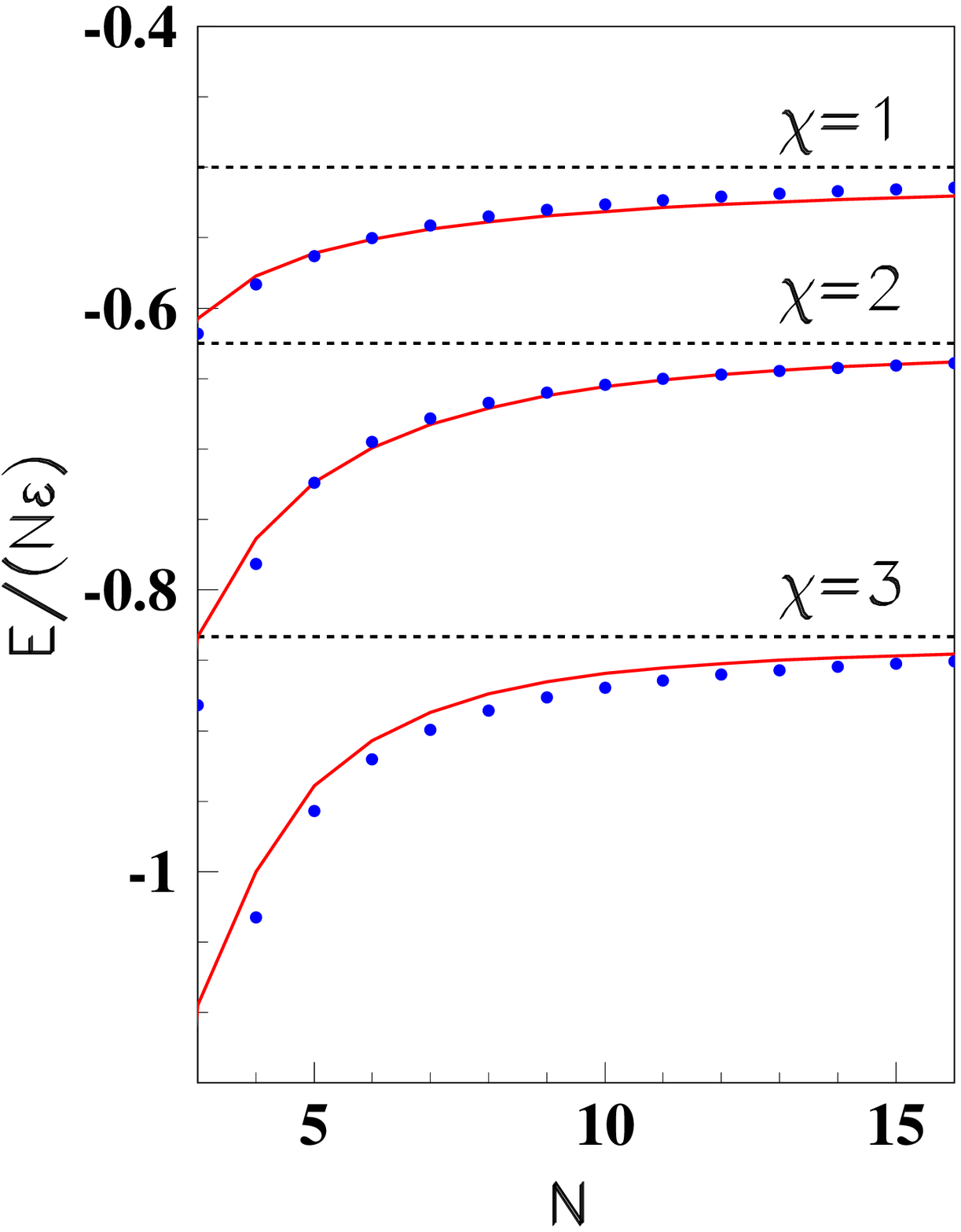}
\caption{Exact ground state energy per particles (solid lines) 
displayed as a function of $N$ for $\chi =1$, $2$ and $3$ from top to bottom. 
In each case, the corresponding HF (dashed line) and DMFT (filled circle) minimum energy 
are shown.
The DMFT calculation is performed using the mean-field and correlation energy resp. given by 
Eq. (\ref{eq:emflipkin}) and Eq. (\ref{eq:coreta}) with $\eta(N) =1.5~N^{-2/3}$.} 
\label{fig:nn}
\end{figure}

The minimal energy  deduced from the semi-empirical density matrix functional proposed above  (filled circles)
is systematically compared to the exact ground state energy (solid line) and HF energy (dashed line) for different 
particle number and two-body interaction strength in Fig. \ref{fig:chi} and \ref{fig:nn}. In all cases, the DMFT 
significantly improves the HF result and turns out to be very close to the exact one. As illustrated 
in Fig. \ref{fig:chi}, the HF energy generally deviates rather significantly from the exact energy around $\chi=1$
and does not provides the correct asymptotic behavior as $\chi$ increases. This deviation, which disappears as 
$N \rightarrow \infty$, is due to the failure of HF theory in the presence of a "shape" phase transition \cite{Aga66}
between the spherical solution ($\chi < 1$) and the "deformed" solution ($\chi > 1$). The standard technique to properly 
account for this effect is to mix different slater determinants like in the GCM theory. 
We see in figure \ref{fig:chi} and \ref{fig:nn} that, except a
small deviation around $\chi=1$ which seems to slightly increase as $N$ increases, both 
the asymptotic behavior at large $\chi$ and $N$  and the energy around $\chi=1$ are rather well reproduced.  
It is worth mentioning that the DMFT is much less demanding in terms of computational power than the 
GCM and thus provides a rather interesting alternative to the latter theory.

\subsection{Discussion on Density Matrix Functional Theory with broken symmetry}
\label{sec:obobs}

\begin{figure}[t!]
\includegraphics[height=9.cm]{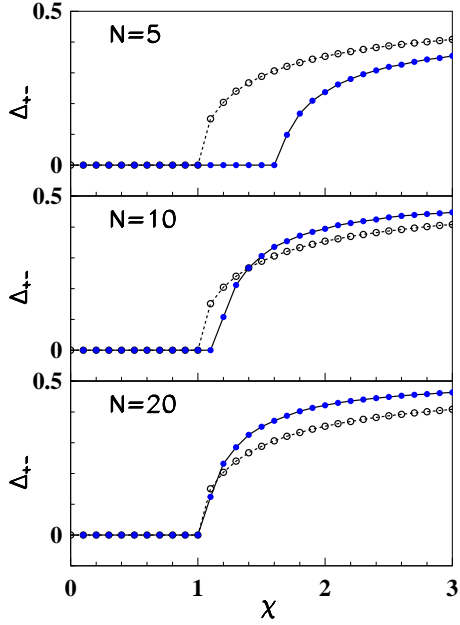}
\caption{Expectation value of $\Delta_{+-}$ as a function of $\chi$ for $N=5$ (top), $N=10$ (middle) and $N=20$ (bottom).  
The DMFT result (filled circles) is systematically compared with the HF (open circles) value.}
\label{fig:ejp}
\end{figure}

Similarly to the Hartree-Fock case, the one-body density solution of the functional developed here 
can eventually violate some symmetries of the "true" ground state density. The
invariance of the Hamiltonian (\ref{eq:hamillipkin}) with respect to parity imposes
$\left\langle  J_+ \right\rangle = \left\langle  J_- \right\rangle = 0$. This implies that at the minimum of
the functional $\alpha=0$. This is indeed the case for the exact functional given by Eq. (\ref{eq:funcn2}) for $N=2$. 
However, solutions with $\alpha \neq 0$ are found for larger particle number. 
This is illustrated in figure \ref{fig:ejp} where the value of 
$\Delta_{+-} \equiv (\left\langle  J_+ \right\rangle + \left\langle  J_- \right\rangle)/2N$ is displayed as a function of $\chi$. 
The  value of $\chi$, for which $\alpha$ becomes non-zero, denoted by $\chi_c$ is infinite for $N=2$, around 1.6 for $N=5$ and tends to the HF case 
($\chi_c = 1$) as $N$ goes to infinity. Note that, in this limit, the HF functional alone already provides a very 
good functional for the Lipkin model.   

\begin{figure}[t!]
\includegraphics[height=9.cm]{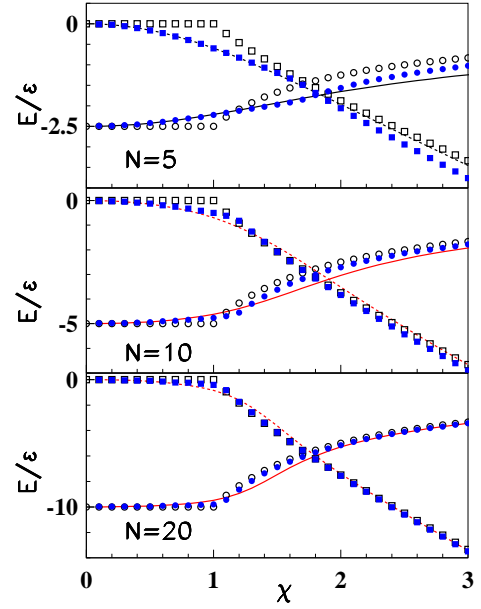}
\caption{Expectation value of the one-body part of the Hamiltonian, denoted by ${\cal E}_{J_0}$ obtained 
at the minimum of the DMFT as a function of $\chi$ for $N=5$ (top), $N=10$ (middle) and $N=20$ (bottom).  
The DMFT result (filled circles) is systematically compared with the exact (solid line) and HF (open circles) value.
The value of the total energy minus ${\cal E}_{J_0}$ is also shown in each panel for the DMFT (filled square), exact (dashed line) and HF 
(open square).} 
\label{fig:ej0}
\end{figure}

According to DMFT, the OBDM $\gamma$ obtained by minimizing the {\it exact} functional should match the exact OBDM at 
the minimum. Therefore, strictly speaking, the 
extracted one-body density and associated natural states and occupation numbers could not be the exact OBDM when some of 
the symmetries of the system are broken. This has  to be kept in mind when discussing expectation values of one-body observables.
As an illustration, the expectation value of the single-particle part of 
the Hamiltonian ${\cal E}_{J_0} \equiv \varepsilon
Tr( \gamma J_0)$ obtained using the OBDM minimizing the DMFT is displayed in Fig. \ref{fig:ej0} as a function of 
$\chi$ for different particle number (filled circles). The exact (solid line) and HF (open circles) prescription are also 
shown.  The value of the total energy minus ${\cal E}_{J_0}$ is also shown in each panel for the DMFT (filled square), 
exact (dashed line) and HF (open square).

The density obtained in DMFT using the semi-empirical functional described in section \ref{sec:semi} always 
improves the estimate of ${\cal E}_{J_0}$ for small $\chi$. It generally gives an almost perfect result when 
$\alpha=0$ at the minimum, i.e. when the parity symmetry is respected by the OBDM.   
The behavior of ${\cal E}_{J_0}$ is also rather satisfactory at large $N$ and $\chi$. In other cases, 
small particle number and large $\chi$ or large particle number and intermediate $\chi$ ($1 \le \chi \le 2$). 
Fig. \ref{fig:ej0} illustrates that, when the functional respects all symmetries of the original Hamiltonian, expectation 
values of one-body observables perfectly match the exact results. This is the case of the semi-empirical functional proposed 
here for all particle number and $\chi \leq 1$. Of course it would be desirable to provide functionals that respects the symmetries 
at the first place. However, as it is well known already at the HF level, the introduction of theories where symmetries are 
explicitly broken is a way to grasp some of the correlations which would have been very hard to incorporate without 
breaking the symmetries. The success of the DMFT functional at large $N$ and $\chi$ can be attributed to the explicit parity
symmetry breaking as in the HF case. 

\section{Summary and discussion on EDF}

In this work, the Density Matrix Functional Theory is applied to the two-level Lipkin model. 
Guided by the $N=2$ case, a semi-empirical functional of the natural wave-functions 
and occupation numbers is constructed. 
The minimization of the DMFT is shown to give a much better agreement  
of the exact ground state energy than the HF scheme over a wide range of particle number 
and two-body interaction strength. 
The success of DMFT in the Lipkin model, shows that this
theory could be a valuable tool in many-body system in the presence of "shape" phase 
transition. Such transition often occurs for instance in nuclear physics \cite{Rin80,Ben03}
and is generally treated by first introducing the  Energy Density Functional of a single-reference 
vacuum (Slater Determinant or quasi-particle state) and then use GCM theory \cite{Ben03}.


DMFT could be a powerful tool to improve actual SR-EDF functionals, by writing directly the EDF in terms of natural 
orbitals and occupation numbers. The possibility to introduce occupation numbers has been promoted in Ref. \cite{Pap07} for the 
pairing Hamiltonian and in \cite{Ber08} for the three-level Lipkin model using a slightly different technique. 
However, the strategy based on DMFT seems quite natural to extend actual EDF. Indeed,   
following the strategy used here for the Lipkin model, we can write the nuclear EDF as   
\begin{eqnarray}
{\cal E}_{EDF}[\{\varphi_i, n_i \} ] = {\cal E}_{MF} [\{\varphi_i, n_i \} ] +  {\cal E}_{C} [\{\varphi_i, n_i \} ] .
\label{eq:edfdmft}
\end{eqnarray}  
The most natural choice for the mean-field part is to use the actual Skyrme functional which has been optimized for decades. 
Note that, the nuclear problem differs from the electronic case and/or the Lipkin model presented here due to the fact 
that coefficients of the functional are not directly linked to the bare interaction but adjusted on experimental data.
Therefore, the mean-field contribution already contains a large fraction of the correlation. Nevertheless,   
the above decomposition (Eq. (\ref{eq:edfdmft})) is already used in the nuclear context in SR-EDF
calculations when a quasi-particle vacuum is retained for the trial state. 
Then, the correlation energy identifies with the pairing energy which could, in the canonical basis, be written as a 
functional of occupation numbers and natural orbitals. Indeed, using the notation $(i,\bar i)$ for canonicals pairs of single-particle states, 
the pairing energy writes
\begin{eqnarray}
{\cal E}_{C} [\{\varphi_i, n_i \} ]  &\equiv& \frac{1}{4} \sum_{i,j} \bar v^{\kappa \kappa}_{i\bar i j \bar j} \sqrt{n_i (1-n_i)} \sqrt{n_j (1-n_j)}
\nonumber \\
\end{eqnarray}
where $\bar v^{\kappa \kappa}$ is the effective interaction in the pairing channels and where we have replaced components of the anomalous density 
$\kappa$  in the natural basis by $\kappa_{i \bar i} = \sqrt{n_i (1-n_i)}$. This functional is adapted for pairing like correlations. 
However, the use of a reference state written as a product of 
quasi-particle states clearly restricts the type of density matrix functional that can be guessed. This class appears 
to be  not general enough to account for the  
diversity of phenomena occurring in nuclei. 
The possibility to use  different functional than the BCS like ones has already been discussed 
in the early time of the Skyrme EDF history \cite{Vau73}. The functional developed here for the Lipkin model 
as well as functionals recently 
proposed in electronic systems \cite{Klo07}, clearly point out the possibility to use
alternative functionals which could be of interest for nuclear systems. 

A crucial aspect of the present work is the introduction of DMFT functionals that explicitly breaks some 
of the symmetries of the original Hamiltonian to incorporate complex correlations. It should be kept in mind 
that broken symmetries imply that symmetries should a priori be restored. The problem of
restoration of broken symmetries in functional theories is an important aspect which deserves 
specific studies in the near future \cite{Lac08,Ben08,Dug08}. 

\begin{acknowledgments}
The author thank M.~Assi\'e, B. Avez, T. Duguet,  C. Simenel, O. Sorlin and P. Van Isacker for enlightening discussions at different stages
of this work and T. Papenbrock for useful remarks on the scaling behavior in the Lipkin model. 

\end{acknowledgments}
   
\bibliography{paper_DMFTlipkin.bbl}

\begin{thebibliography}{42}
\expandafter\ifx\csname natexlab\endcsname\relax\def\natexlab#1{#1}\fi
\expandafter\ifx\csname bibnamefont\endcsname\relax
  \def\bibnamefont#1{#1}\fi
\expandafter\ifx\csname bibfnamefont\endcsname\relax
  \def\bibfnamefont#1{#1}\fi
\expandafter\ifx\csname citenamefont\endcsname\relax
  \def\citenamefont#1{#1}\fi
\expandafter\ifx\csname url\endcsname\relax
  \def\url#1{\texttt{#1}}\fi
\expandafter\ifx\csname urlprefix\endcsname\relax\def\urlprefix{URL }\fi
\providecommand{\bibinfo}[2]{#2}
\providecommand{\eprint}[2][]{\url{#2}}

\bibitem[{\citenamefont{Bender et~al.}(2003)\citenamefont{Bender, Heenen, and
  Reinhard}}]{Ben03}
\bibinfo{author}{\bibfnamefont{M.}~\bibnamefont{Bender}},
  \bibinfo{author}{\bibfnamefont{P.-H.} \bibnamefont{Heenen}},
  \bibnamefont{and} \bibinfo{author}{\bibfnamefont{P.-G.}
  \bibnamefont{Reinhard}}, \bibinfo{journal}{Rev. Mod. Phys.}
  \textbf{\bibinfo{volume}{75}}, \bibinfo{pages}{121} (\bibinfo{year}{2003}).

\bibitem[{\citenamefont{Stone and Reinhard}(2007)}]{Sto07}
\bibinfo{author}{\bibfnamefont{J.}~\bibnamefont{Stone}} \bibnamefont{and}
  \bibinfo{author}{\bibfnamefont{P.-G.} \bibnamefont{Reinhard}},
  \bibinfo{journal}{Prog. Part. and Nucl. Phys.} \textbf{\bibinfo{volume}{58}},
  \bibinfo{pages}{587} (\bibinfo{year}{2007}).

\bibitem[{\citenamefont{Ring and Schuck}(1980)}]{Rin80}
\bibinfo{author}{\bibfnamefont{P.}~\bibnamefont{Ring}} \bibnamefont{and}
  \bibinfo{author}{\bibfnamefont{P.}~\bibnamefont{Schuck}},
  \emph{\bibinfo{title}{The Nuclear Many-Body Problem}}
  (\bibinfo{publisher}{Springer-Verlag}, \bibinfo{address}{New-York},
  \bibinfo{year}{1980}).

\bibitem[{\citenamefont{Dobaczewski et~al.}(2007)\citenamefont{Dobaczewski,
  Stoitsov, Nazarewicz, and Reinhard}}]{Dob07}
\bibinfo{author}{\bibfnamefont{J.}~\bibnamefont{Dobaczewski}},
  \bibinfo{author}{\bibfnamefont{M.~V.} \bibnamefont{Stoitsov}},
  \bibinfo{author}{\bibfnamefont{W.}~\bibnamefont{Nazarewicz}},
  \bibnamefont{and} \bibinfo{author}{\bibfnamefont{P.-G.}
  \bibnamefont{Reinhard}}, \bibinfo{journal}{Physical Review C (Nuclear
  Physics)} \textbf{\bibinfo{volume}{76}}, \bibinfo{eid}{054315}
  (pages~\bibinfo{numpages}{16}) (\bibinfo{year}{2007}),
  \urlprefix\url{http://link.aps.org/abstract/PRC/v76/e054315}.

\bibitem[{\citenamefont{Lacroix et~al.}(2008)\citenamefont{Lacroix, Duguet, and
  Bender}}]{Lac08}
\bibinfo{author}{\bibfnamefont{D.}~\bibnamefont{Lacroix}},
  \bibinfo{author}{\bibfnamefont{T.}~\bibnamefont{Duguet}}, \bibnamefont{and}
  \bibinfo{author}{\bibfnamefont{M.}~\bibnamefont{Bender}},
  \bibinfo{journal}{Phys. Rev. C} \textbf{\bibinfo{volume}{xxx}},
  \bibinfo{pages}{xxx} (\bibinfo{year}{2008}), \bibinfo{note}{in preparation}.

\bibitem[{\citenamefont{Bender et~al.}(2008)\citenamefont{Bender, Duguet, and
  Lacroix}}]{Ben08}
\bibinfo{author}{\bibfnamefont{M.}~\bibnamefont{Bender}},
  \bibinfo{author}{\bibfnamefont{T.}~\bibnamefont{Duguet}}, \bibnamefont{and}
  \bibinfo{author}{\bibfnamefont{D.}~\bibnamefont{Lacroix}},
  \bibinfo{journal}{Phys. Rev. C} \textbf{\bibinfo{volume}{xxx}},
  \bibinfo{pages}{xxx} (\bibinfo{year}{2008}), \bibinfo{note}{in preparation}.

\bibitem[{\citenamefont{Duguet et~al.}(2008)\citenamefont{Duguet, Bender,
  Bennaceur, Lacroix, and Lesinski}}]{Dug08}
\bibinfo{author}{\bibfnamefont{T.}~\bibnamefont{Duguet}},
  \bibinfo{author}{\bibfnamefont{M.}~\bibnamefont{Bender}},
  \bibinfo{author}{\bibfnamefont{K.}~\bibnamefont{Bennaceur}},
  \bibinfo{author}{\bibfnamefont{D.}~\bibnamefont{Lacroix}}, \bibnamefont{and}
  \bibinfo{author}{\bibfnamefont{T.}~\bibnamefont{Lesinski}},
  \bibinfo{journal}{Phys. Rev. C} \textbf{\bibinfo{volume}{xxx}},
  \bibinfo{pages}{xxx} (\bibinfo{year}{2008}), \bibinfo{note}{in preparation}.

\bibitem[{\citenamefont{Gilbert}(1975)}]{Gil75}
\bibinfo{author}{\bibfnamefont{T.~L.} \bibnamefont{Gilbert}},
  \bibinfo{journal}{Phys. Rev. B} \textbf{\bibinfo{volume}{12}},
  \bibinfo{pages}{2111} (\bibinfo{year}{1975}).

\bibitem[{\citenamefont{Goedecker and Umrigar}(2000)}]{Umr00}
\bibinfo{author}{\bibfnamefont{S.}~\bibnamefont{Goedecker}} \bibnamefont{and}
  \bibinfo{author}{\bibfnamefont{C.}~\bibnamefont{Umrigar}},
  \emph{\bibinfo{title}{Natural Orbital Functional Theory}}
  (\bibinfo{publisher}{Kluwer Academic/Plenum Publishers, Ed. J. Ciolowski},
  \bibinfo{address}{New-York}, \bibinfo{year}{2000}).

\bibitem[{\citenamefont{Kollmar}(2006)}]{Kol06}
\bibinfo{author}{\bibfnamefont{C.}~\bibnamefont{Kollmar}},
  \bibinfo{journal}{The Journal of Chemical Physics}
  \textbf{\bibinfo{volume}{125}}, \bibinfo{eid}{084108}
  (pages~\bibinfo{numpages}{12}) (\bibinfo{year}{2006}).

\bibitem[{\citenamefont{H.~J.~Lipkin}(1965)}]{Lip65}
\bibinfo{author}{\bibfnamefont{A.~J.~G.} \bibnamefont{H.~J.~Lipkin},
  \bibfnamefont{N.~Meshkov}}, \bibinfo{journal}{Nucl. Phys. A}
  \textbf{\bibinfo{volume}{62}}, \bibinfo{pages}{188} (\bibinfo{year}{1965}).

\bibitem[{\citenamefont{D.~Agassi}(1966)}]{Aga66}
\bibinfo{author}{\bibfnamefont{N.~M.} \bibnamefont{D.~Agassi},
  \bibfnamefont{H.~J.~Lipkin}}, \bibinfo{journal}{Nucl. Phys. A}
  \textbf{\bibinfo{volume}{86}}, \bibinfo{pages}{321} (\bibinfo{year}{1966}).

\bibitem[{\citenamefont{Hohenberg and Kohn}(1964)}]{Hoh64}
\bibinfo{author}{\bibfnamefont{P.}~\bibnamefont{Hohenberg}} \bibnamefont{and}
  \bibinfo{author}{\bibfnamefont{W.}~\bibnamefont{Kohn}},
  \bibinfo{journal}{Phys. Rev.} \textbf{\bibinfo{volume}{136}},
  \bibinfo{pages}{B864} (\bibinfo{year}{1964}).

\bibitem[{\citenamefont{Kohn and Sham}(1965)}]{Koh65}
\bibinfo{author}{\bibfnamefont{W.}~\bibnamefont{Kohn}} \bibnamefont{and}
  \bibinfo{author}{\bibfnamefont{L.~J.} \bibnamefont{Sham}},
  \bibinfo{journal}{Phys. Rev.} \textbf{\bibinfo{volume}{140}},
  \bibinfo{pages}{A1133} (\bibinfo{year}{1965}).

\bibitem[{\citenamefont{Valone}(1980{\natexlab{a}})}]{Val80a}
\bibinfo{author}{\bibfnamefont{S.~M.} \bibnamefont{Valone}},
  \bibinfo{journal}{The Journal of Chemical Physics}
  \textbf{\bibinfo{volume}{73}}, \bibinfo{pages}{1344}
  (\bibinfo{year}{1980}{\natexlab{a}}),
  \urlprefix\url{http://link.aip.org/link/?JCP/73/1344/1}.

\bibitem[{\citenamefont{Valone}(1980{\natexlab{b}})}]{Val80b}
\bibinfo{author}{\bibfnamefont{S.~M.} \bibnamefont{Valone}},
  \bibinfo{journal}{The Journal of Chemical Physics}
  \textbf{\bibinfo{volume}{73}}, \bibinfo{pages}{4653}
  (\bibinfo{year}{1980}{\natexlab{b}}),
  \urlprefix\url{http://link.aip.org/link/?JCP/73/4653/1}.

\bibitem[{\citenamefont{Zumbach and Maschke}(1985)}]{Zum85}
\bibinfo{author}{\bibfnamefont{G.}~\bibnamefont{Zumbach}} \bibnamefont{and}
  \bibinfo{author}{\bibfnamefont{K.}~\bibnamefont{Maschke}},
  \bibinfo{journal}{The Journal of Chemical Physics}
  \textbf{\bibinfo{volume}{82}}, \bibinfo{pages}{5604} (\bibinfo{year}{1985}),
  \urlprefix\url{http://link.aip.org/link/?JCP/82/5604/1}.

\bibitem[{\citenamefont{Muller}(1984)}]{Mul84}
\bibinfo{author}{\bibfnamefont{A.}~\bibnamefont{Muller}},
  \bibinfo{journal}{Phys. Lett. A} \textbf{\bibinfo{volume}{105}},
  \bibinfo{pages}{446} (\bibinfo{year}{1984}).

\bibitem[{\citenamefont{Goedecker and Umrigar}(1998)}]{Goe98}
\bibinfo{author}{\bibfnamefont{S.}~\bibnamefont{Goedecker}} \bibnamefont{and}
  \bibinfo{author}{\bibfnamefont{C.~J.} \bibnamefont{Umrigar}},
  \bibinfo{journal}{Phys. Rev. Lett.} \textbf{\bibinfo{volume}{81}},
  \bibinfo{pages}{866} (\bibinfo{year}{1998}).

\bibitem[{\citenamefont{Cs\'anyi and Arias}(2000)}]{Csa00}
\bibinfo{author}{\bibfnamefont{G.}~\bibnamefont{Cs\'anyi}} \bibnamefont{and}
  \bibinfo{author}{\bibfnamefont{T.~A.} \bibnamefont{Arias}},
  \bibinfo{journal}{Phys. Rev. B} \textbf{\bibinfo{volume}{61}},
  \bibinfo{pages}{7348} (\bibinfo{year}{2000}).

\bibitem[{\citenamefont{Cs\'anyi et~al.}(2002)\citenamefont{Cs\'anyi,
  Goedecker, and Arias}}]{Csa02}
\bibinfo{author}{\bibfnamefont{G.}~\bibnamefont{Cs\'anyi}},
  \bibinfo{author}{\bibfnamefont{S.}~\bibnamefont{Goedecker}},
  \bibnamefont{and} \bibinfo{author}{\bibfnamefont{T.~A.} \bibnamefont{Arias}},
  \bibinfo{journal}{Phys. Rev. A} \textbf{\bibinfo{volume}{65}},
  \bibinfo{pages}{032510} (\bibinfo{year}{2002}).

\bibitem[{\citenamefont{Yasuda}(2002)}]{Yas02}
\bibinfo{author}{\bibfnamefont{K.}~\bibnamefont{Yasuda}},
  \bibinfo{journal}{Phys. Rev. Lett.} \textbf{\bibinfo{volume}{88}},
  \bibinfo{pages}{053001} (\bibinfo{year}{2002}).

\bibitem[{\citenamefont{Kollmar}(2004)}]{Kol04}
\bibinfo{author}{\bibfnamefont{C.}~\bibnamefont{Kollmar}},
  \bibinfo{journal}{The Journal of Chemical Physics}
  \textbf{\bibinfo{volume}{121}}, \bibinfo{pages}{11581}
  (\bibinfo{year}{2004}),
  \urlprefix\url{http://link.aip.org/link/?JCP/121/11581/1}.

\bibitem[{\citenamefont{Cioslowski et~al.}(2003)\citenamefont{Cioslowski,
  Pernal, and Buchowiecki}}]{Cio03}
\bibinfo{author}{\bibfnamefont{J.}~\bibnamefont{Cioslowski}},
  \bibinfo{author}{\bibfnamefont{K.}~\bibnamefont{Pernal}}, \bibnamefont{and}
  \bibinfo{author}{\bibfnamefont{M.}~\bibnamefont{Buchowiecki}},
  \bibinfo{journal}{The Journal of Chemical Physics}
  \textbf{\bibinfo{volume}{119}}, \bibinfo{pages}{6443} (\bibinfo{year}{2003}),
  \urlprefix\url{http://link.aip.org/link/?JCP/119/6443/1}.

\bibitem[{\citenamefont{Pernal and Cioslowski}(2004)}]{Per04}
\bibinfo{author}{\bibfnamefont{K.}~\bibnamefont{Pernal}} \bibnamefont{and}
  \bibinfo{author}{\bibfnamefont{J.}~\bibnamefont{Cioslowski}},
  \bibinfo{journal}{The Journal of Chemical Physics}
  \textbf{\bibinfo{volume}{120}}, \bibinfo{pages}{5987} (\bibinfo{year}{2004}),
  \urlprefix\url{http://link.aip.org/link/?JCP/120/5987/1}.

\bibitem[{\citenamefont{Gritsenko et~al.}(2005)\citenamefont{Gritsenko, Pernal,
  and Baerends}}]{Gri05}
\bibinfo{author}{\bibfnamefont{O.}~\bibnamefont{Gritsenko}},
  \bibinfo{author}{\bibfnamefont{K.}~\bibnamefont{Pernal}}, \bibnamefont{and}
  \bibinfo{author}{\bibfnamefont{E.~J.} \bibnamefont{Baerends}},
  \bibinfo{journal}{The Journal of Chemical Physics}
  \textbf{\bibinfo{volume}{122}}, \bibinfo{eid}{204102}
  (pages~\bibinfo{numpages}{13}) (\bibinfo{year}{2005}),
  \urlprefix\url{http://link.aip.org/link/?JCP/122/204102/1}.

\bibitem[{\citenamefont{Lathiotakis et~al.}(2005)\citenamefont{Lathiotakis,
  Helbig, and Gross}}]{Lat05}
\bibinfo{author}{\bibfnamefont{N.~N.} \bibnamefont{Lathiotakis}},
  \bibinfo{author}{\bibfnamefont{N.}~\bibnamefont{Helbig}}, \bibnamefont{and}
  \bibinfo{author}{\bibfnamefont{E.~K.~U.} \bibnamefont{Gross}},
  \bibinfo{journal}{Physical Review A (Atomic, Molecular, and Optical Physics)}
  \textbf{\bibinfo{volume}{72}}, \bibinfo{eid}{030501}
  (pages~\bibinfo{numpages}{4}) (\bibinfo{year}{2005}),
  \urlprefix\url{http://link.aps.org/abstract/PRA/v72/e030501}.

\bibitem[{\citenamefont{Cioslowski and Pernal}(2005)}]{Cio05}
\bibinfo{author}{\bibfnamefont{J.}~\bibnamefont{Cioslowski}} \bibnamefont{and}
  \bibinfo{author}{\bibfnamefont{K.}~\bibnamefont{Pernal}},
  \bibinfo{journal}{Physical Review B (Condensed Matter and Materials Physics)}
  \textbf{\bibinfo{volume}{71}}, \bibinfo{eid}{113103}
  (pages~\bibinfo{numpages}{4}) (\bibinfo{year}{2005}).

\bibitem[{\citenamefont{Leiva and Piris}(2005)}]{Lei05}
\bibinfo{author}{\bibfnamefont{P.}~\bibnamefont{Leiva}} \bibnamefont{and}
  \bibinfo{author}{\bibfnamefont{M.}~\bibnamefont{Piris}},
  \bibinfo{journal}{The Journal of Chemical Physics}
  \textbf{\bibinfo{volume}{123}}, \bibinfo{eid}{214102}
  (pages~\bibinfo{numpages}{7}) (\bibinfo{year}{2005}).

\bibitem[{\citenamefont{Marques and Lathiotakis}(2008)}]{Mar08}
\bibinfo{author}{\bibfnamefont{M.~A.~L.} \bibnamefont{Marques}}
  \bibnamefont{and} \bibinfo{author}{\bibfnamefont{N.~N.}
  \bibnamefont{Lathiotakis}}, \bibinfo{journal}{Physical Review A (Atomic,
  Molecular, and Optical Physics)} \textbf{\bibinfo{volume}{77}},
  \bibinfo{eid}{032509} (pages~\bibinfo{numpages}{4}) (\bibinfo{year}{2008}),
  \urlprefix\url{http://link.aps.org/abstract/PRA/v77/e032509}.

\bibitem[{\citenamefont{Lathiotakis and Marques}(2008)}]{Lat08}
\bibinfo{author}{\bibfnamefont{N.~N.} \bibnamefont{Lathiotakis}}
  \bibnamefont{and} \bibinfo{author}{\bibfnamefont{M.~A.~L.}
  \bibnamefont{Marques}}, \emph{\bibinfo{title}{Benchmark calculations for
  reduced density-matrix functional theory}} (\bibinfo{year}{2008}),
  \urlprefix\url{http://www.citebase.org/abstract?id=oai:arXiv.org:0801.3439}.

\bibitem[{\citenamefont{Cioslowski and Pernal}(1999)}]{Cio99}
\bibinfo{author}{\bibfnamefont{J.}~\bibnamefont{Cioslowski}} \bibnamefont{and}
  \bibinfo{author}{\bibfnamefont{K.}~\bibnamefont{Pernal}},
  \bibinfo{journal}{J. Chem. Phys.} \textbf{\bibinfo{volume}{111}},
  \bibinfo{pages}{3396} (\bibinfo{year}{1999}).

\bibitem[{\citenamefont{Lathiotakis et~al.}(2007)\citenamefont{Lathiotakis,
  Helbig, and Gross}}]{Lat07}
\bibinfo{author}{\bibfnamefont{N.~N.} \bibnamefont{Lathiotakis}},
  \bibinfo{author}{\bibfnamefont{N.}~\bibnamefont{Helbig}}, \bibnamefont{and}
  \bibinfo{author}{\bibfnamefont{E.~K.~U.} \bibnamefont{Gross}},
  \bibinfo{journal}{Physical Review B (Condensed Matter and Materials Physics)}
  \textbf{\bibinfo{volume}{75}}, \bibinfo{eid}{195120}
  (pages~\bibinfo{numpages}{8}) (\bibinfo{year}{2007}),
  \urlprefix\url{http://link.aps.org/abstract/PRB/v75/e195120}.

\bibitem[{\citenamefont{Severyukhin et~al.}(2006)\citenamefont{Severyukhin,
  Bender, and Heenen}}]{Sev06}
\bibinfo{author}{\bibfnamefont{A.~P.} \bibnamefont{Severyukhin}},
  \bibinfo{author}{\bibfnamefont{M.}~\bibnamefont{Bender}}, \bibnamefont{and}
  \bibinfo{author}{\bibfnamefont{P.-H.} \bibnamefont{Heenen}},
  \bibinfo{journal}{Physical Review C (Nuclear Physics)}
  \textbf{\bibinfo{volume}{74}}, \bibinfo{eid}{024311}
  (pages~\bibinfo{numpages}{7}) (\bibinfo{year}{2006}),
  \urlprefix\url{http://link.aps.org/abstract/PRC/v74/e024311}.

\bibitem[{\citenamefont{Ciolowski}(2000)}]{Cio00}
\bibinfo{author}{\bibfnamefont{J.}~\bibnamefont{Ciolowski}},
  \emph{\bibinfo{title}{Many-Electron densities and reduced density matrices}}
  (\bibinfo{publisher}{Kluwer Academic/Plenum Publishers},
  \bibinfo{address}{New-York}, \bibinfo{year}{2000}).

\bibitem[{\citenamefont{Lacroix et~al.}(2004)\citenamefont{Lacroix, Ayik, and
  Chomaz}}]{Lac04}
\bibinfo{author}{\bibfnamefont{D.}~\bibnamefont{Lacroix}},
  \bibinfo{author}{\bibfnamefont{S.}~\bibnamefont{Ayik}}, \bibnamefont{and}
  \bibinfo{author}{\bibfnamefont{P.}~\bibnamefont{Chomaz}},
  \bibinfo{journal}{Progress in Part. and Nucl. Phys.}
  \textbf{\bibinfo{volume}{52}}, \bibinfo{pages}{497} (\bibinfo{year}{2004}).

\bibitem[{\citenamefont{Klooster}(2007)}]{Klo07}
\bibinfo{author}{\bibfnamefont{R.}~\bibnamefont{Klooster}},
  \emph{\bibinfo{title}{Density matrix functional theory}}
  (\bibinfo{year}{2007}),
  \urlprefix\url{http://theochem.chem.rug.nl/publications/Abstracts.html#587}.

\bibitem[{\citenamefont{L\"owdin and Shull}(1956)}]{Low56}
\bibinfo{author}{\bibfnamefont{P.-O.} \bibnamefont{L\"owdin}} \bibnamefont{and}
  \bibinfo{author}{\bibfnamefont{H.}~\bibnamefont{Shull}},
  \bibinfo{journal}{Phys. Rev.} \textbf{\bibinfo{volume}{101}},
  \bibinfo{pages}{1730} (\bibinfo{year}{1956}).

\bibitem[{\citenamefont{Dusuel and Vidal}(2004)}]{Dus04}
\bibinfo{author}{\bibfnamefont{S.}~\bibnamefont{Dusuel}} \bibnamefont{and}
  \bibinfo{author}{\bibfnamefont{J.}~\bibnamefont{Vidal}},
  \bibinfo{journal}{Physical Review Letters} \textbf{\bibinfo{volume}{93}},
  \bibinfo{eid}{237204} (pages~\bibinfo{numpages}{4}) (\bibinfo{year}{2004}),
  \urlprefix\url{http://link.aps.org/abstract/PRL/v93/e237204}.

\bibitem[{\citenamefont{Papenbrock and Bhattacharyya}(2007)}]{Pap07}
\bibinfo{author}{\bibfnamefont{T.}~\bibnamefont{Papenbrock}} \bibnamefont{and}
  \bibinfo{author}{\bibfnamefont{A.}~\bibnamefont{Bhattacharyya}},
  \bibinfo{journal}{Physical Review C (Nuclear Physics)}
  \textbf{\bibinfo{volume}{75}}, \bibinfo{eid}{014304}
  (pages~\bibinfo{numpages}{7}) (\bibinfo{year}{2007}),
  \urlprefix\url{http://link.aps.org/abstract/PRC/v75/e014304}.

\bibitem[{\citenamefont{Bertolli and Papenbrock}(2008)}]{Ber08}
\bibinfo{author}{\bibfnamefont{M.~G.} \bibnamefont{Bertolli}} \bibnamefont{and}
  \bibinfo{author}{\bibfnamefont{T.}~\bibnamefont{Papenbrock}}
  (\bibinfo{year}{2008}), \eprint{0805.2856}.

\bibitem[{\citenamefont{Vautherin}(1973)}]{Vau73}
\bibinfo{author}{\bibfnamefont{D.}~\bibnamefont{Vautherin}},
  \bibinfo{journal}{Phys. Rev. C} \textbf{\bibinfo{volume}{7}},
  \bibinfo{pages}{296} (\bibinfo{year}{1973}).

\end{thebibliography}
\end{document}